\documentclass[twocolumn,english,aps,prl,showpacs]{revtex4}
\usepackage{graphicx}

\makeatletter

\usepackage{babel}
\makeatother
\begin{document}

\title{NonAbelian states with negative flux~: a new series of quantum Hall states}

\author{Th.~Jolicoeur}

\affiliation{Laboratoire de Physique Th\'eorique et Mod\`eles Statistiques, 
Universit\'e Paris-Sud, 91405 Orsay Cedex, France}


\begin{abstract}
By applying the idea of parafermionic clustering to composite bosons
with positive as well as \textit{negative} flux, a new series of trial wavefunctions
to describe fractional quantum Hall states is proposed. These states compete
at filling factors $\nu =k/(3k\pm 2)$ with other ground states like stripes or
composite fermion states.
These series contain all the states recently discovered by Pan et al. 
[Phys. Rev. Lett. \textbf{90}, 016801 (2003)] including the even denominator cases.
Exact diagonalization studies on the sphere and torus point to their relevance for
 $\nu =3/7, 3/11$, and $3/8$.
\end{abstract}


\maketitle


Two-dimensional electron gases in a quantizing magnetic field display a wealth
of incompressible liquid phases at low temperature. These liquids may be
classified by sequences of special values of Landau level (LL) filling factors $\nu$.
The most prominent sequences are observed for $\nu =p/(2mp\pm 1)$, $p$ and $m$ integers.
In the composite-fermion picture~\cite{Jain89,Lopez91,Kalmeyer92,DasSarma,Heinonen}, 
this is interpreted as an integer Hall effect
for composites made of an electron bound to $2m$ flux quanta, the composite fermions
$^{2m}$CFs.

Recent experiments~\cite{Pan03} have uncovered states displaying the fractional quantum Hall effect
(FQHE)  at filling factors
$\nu =4/11, 5/13, 4/13, 6/17$, and $5/17$ that do not belong to the primary
FQHE sequences. In addition, there is also evidence for two even-denominator fractions
 $\nu =3/10$, and $3/8$. This is very unusual since the only previously known example
of an even-denominator fraction is the elusive $\nu =5/2$ state. The state $3/8$ has 
also been observed~\cite{Xia04} in the N=1 LL at total filling factor $\nu = 2+3/8$.
The new odd-denominator fractions can be explained by hierarchical reasoning
in the spirit of the original Halperin-Haldane hierarchy. For example, at $\nu =4/11$, the 
$^2$CFs have an effective filling factor $\nu_{CF} = 1+ 1/3$. 
If the interactions between the $^2$CFs
have a repulsive short-range core then it is plausible that they will themselves form a
standard Laughlin liquid at filling factor 1/3 within the second CF Landau level.
This possibility pointed out in the work of Pan et al.~\cite{Pan03} has been explored 
theoretically~\cite{Lopez04,Goerbig04,Chang04}. It should be pointed out that this construction of
``second generation'' of composite fermions is part of the 
standard lore of the
hierarchical view of the FQHE states since the CF construction and the older Halperin-Haldane
hierarchy can be related by a change of basis in the lattice of quantum numbers~\cite{Read90}.
Since the even-denominator fractions requires clustering they do not fit naturally in this picture.

In this Letter, I propose a construction based on the idea of composite bosons that carry now
an odd number of flux quanta. These fluxes may be positive or negative. I then exploit the 
possibility of clustering of bosons in the lowest LL (LLL). Indeed it has been suggested~\cite{Cooper01} 
that incompressible liquids of Bose particle may form at fillings $\nu =k/2$ with integer $k$. 
I write down 
spin-polarized FQHE wavefunctions on the disk and spherical geometry. By construction they reside
entirely in the LLL and have filling factor $\nu = k/(3k\pm 2)$. 
All these states are non-Abelian FQHE states with unconventional excitations.
While the positive flux series
 already appeared in
the work of Read and Rezayi~\cite{Read96,Read99}, the negative flux series is new. 
These series produce candidate wavefunctions for all the states observed by Pan et al. beyond
the main CF sequences, thus unifying even and odd denominator fractions.
For the fraction 3/7, the negative flux
candidate wavefunction has an excellent overlap with the Coulomb ground state obtained by exact
diagonalization on the sphere for N=6 electrons. There is also a two-fold possible topological 
degeneracy on the torus geometry obersevd by tweaking the Coulomb interaction.
The possible non-Abelian 3/11 state is also observed on the sphere. Finally
at $\nu =3/8$ there is a seven-fold quasi degeneracy of the ground state on the torus
also compatible with a non-Abelian state.

The first observation is that some of the new fractions of ref.(\cite{Pan03}) are of the form
$p/(3p\pm 1)$. This would be natural for the FQHE of \textit{bosons} where one 
expects the formation
of composite fermions with an odd number of flux tubes, i.e. $^1$CF and $^3$CF. 
The $^1$CF lead to a series of Bose fractions at $\nu = p/(p+1)$ which has nothing to do with
the present problem.
But if the $^3$CFs fill an integer number of pseudo-Landau levels then this leads
to magic fillings $p/(3p\pm 1)$. Indeed there is evidence
from theoretical studies of bosons in the LLL with dipolar interactions~\cite{Rezayi05}
 that such $^3$CF do appear.
This suggests that composite bosons may form in the electronic system, three flux tubes bound 
to one electron, $^3$CBs, the attachment may be with statistical flux along or against the 
applied magnetic field. If $\nu$ stands for the electron filling factor and $\nu^*$ the 
$^3$CB filling factor, 
they are related by $1/\nu = 3 +1/\nu^*$. The relationship 
between the wanted electronic trial wavefunction and the CB wavefunction is~:
\begin{equation}
\Psi^{Fermi}_\nu (\{z_i\})=
\mathcal{P}_{LLL}\prod_{i<j}(z_i-z_j)^3\,\, \Phi^{Bose}_{\nu^*}(\{z_i,z_i^*\}),
 \label{wavef}
\end{equation} 
where $z_i=x_i\, +\, iy_i$ refer to the coordinates of the electrons 
in the unbounded disk geometry and the symmetric gauge,
$\mathcal{P}_{LLL}$
is the LLL projection operator. 
The Laughlin-Jastrow factor $\prod_{i<j}(z_i-z_i)^3$ transforms bosons into fermions and
adequately takes into account the Coulomb repulsion.
The next step is to find candidates for the trial state $\Phi^{Bose}_{\nu^*}$.
It has been suggested~\cite{Cooper01} that bosons in the LLL may form incompressible
states for $\nu^*=k/2$. There is evidence that they are described by
 the Read-Rezayi parafermionic states~\cite{Read96,Read99} with clustering of $k$
particles~:
\begin{equation}
 \Phi^{RR}_{\nu^*=k/2}=\mathcal{S}\,\,
\left[
\prod_{i_1<j_1}(z_{i_1}-z_{j_1})^2 \dots
\prod_{i_k<j_k}(z_{i_k}-z_{j_k})^2 
\right].
\label{RR}
\end{equation} 
In this equation, the $\mathcal{S}$ symbol means  symmetrization of the product
of Laughlin-Jastrow factors over all partition of N particles in subsets of $N/k$
particles ($N$ being divisible by $k$). 
The ubiquitous exponential factor appearing in all LLL states has been omitted for clarity.
While the relevance of such states to bosons
with contact interactions is not clear, it has been shown that longer-range interactions 
like dipolar
interaction may help stabilize these states~\cite{Rezayi05}. Since the CBs are composite objects
it is likely that their mutual interaction has also some long-range character. It is
thus natural to try the ansatz $\Phi^{Bose}_{\nu^*}=\Phi^{RR}_{\nu^*=k/2}$ in Eq.(\ref{wavef}).
This leads to a series of states with electron filling factor $\nu =k/(3k+2)$ which is in fact
the $M=3$ case of the generalized $(k,M)$ states constructed by Read and Rezayi. In this construction, 
the flux attached to the boson is positive. It is also possible to construct wavefunctions
with negative flux~\cite{Moller05} attached to the CBs. Now the Bose function depens only upon
the antiholomorphic coordinates~:
\begin{equation}
\Phi^{Bose}_{\nu^*}(\{z_i^*\})=(\Phi^{RR}_{\nu^*=k/2}(\{z_i\}))^*
 \label{negflux}
\end{equation} 
The projection onto the LLL in Eq.(\ref{wavef}) means that the electronic wavefunction
can be written as~:
\begin{equation}
\Psi^{Fermi}_\nu (\{z_i\})= \Phi^{RR}_{\nu^*=k/2}(\{\frac{\partial}{\partial z_i}\})
\prod_{i<j}(z_i-z_j)^3 .
 \label{NGwavef}
\end{equation} 
The filling factor of this new series of states is now $\nu = k/(3k-2)$.
These states can be written in the spherical geometry with the help of the spinor components
$u_i=\cos (\theta_i/2) {\rm e}^{i\varphi_i/2}$,
$v_i=\sin (\theta_i/2) {\rm e}^{-i\varphi_i/2}$ ($\{\theta_i,\phi_i\} $ being standard polar coordinates) 
by making the following substitutions~:
\begin{equation}
z_i-z_j \rightarrow u_iv_j - u_j v_i ,
\quad \partial_{z_i}-\partial_{z_j}
\rightarrow 
\partial_{u_i}\partial_{v_j} -
\partial_{v_i}\partial_{u_j}.
\end{equation} 
This construction leads to wavefunctions that have zero total angular momentum $L=0$
as expected for liquid states.
On the sphere the two series of states have a definite relation between the number
of flux quanta through the surface and the number of electrons. The positive flux series
has $N_\phi = N/\nu - 5$ while the negative flux series has $N_\phi = N/\nu - 1$.
Even when these states have the same filling factor as standard hierarchy/composite fermion states,
the shift (the constant term in the $N_\phi - N$ relation) is in general different.
The positive flux series starts with the Laughlin state for $\nu =1/5$ at $k=1$,
the $k=2$ state is the known Pfaffian state~\cite{Moore91,Greiter92} at $\nu=1/4$,
at $k=3$ there a state with $\nu =3/11$ which compete with the $^4$CF state with negative flux,
at $k=4$ the competition is with the similar $\nu =2/7$  $^4$CF state. This series also contains
5/17 at $k=5$, 3/10 at $k=6$, and 4/13 at $k=8$. The negative flux series starts with the filled
Landau level at $k=1$ and contains notably 5/13 ($k=5$), 3/8 ($k=6$), 4/11 ($k=8$), 6/17 ($k=12$).
It is not likely that these states will compete favorably with the main sequence CF states
in view of their remarkable stability. However the situation is open concerning the exotic even
 denominator and the unconventional odd-denominator states. Also the CF states may be destabilized
by tuning the interaction potential. A two-body interaction in a given LL may always
be parameterized by the pseudopotentials $V_m$, $m=1,3,\dots $,where $V_m$ is the 
interaction energy for a single pair of electrons with relative angular momentum $m$
(all energies will be expressed in units of $e^2 /\epsilon l_0 $ and $l_0=\sqrt{\hbar c/eB} $).
It is known for example that there is a window of stability
for a non-Abelian $\nu =2/5$ state in the N=1 LL~\cite{Rezayi06} which is obtained by
slightly decreasing the $V_1$ component with respect to its Coulomb value.

\begin{figure}
  \begin{center}
    \begin{tabular}{cc}
      \resizebox{45mm}{!}{\includegraphics{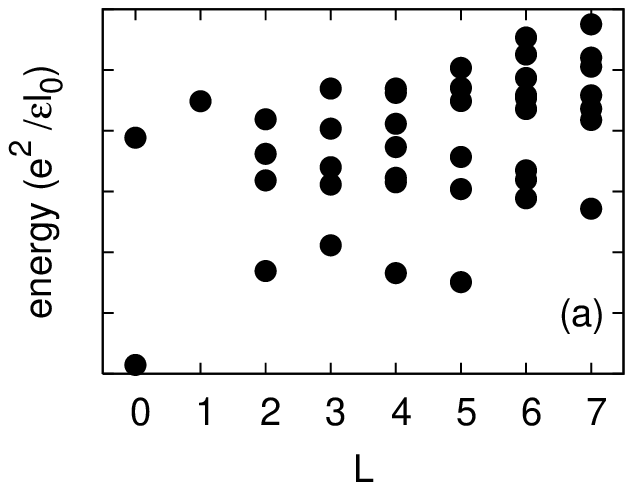}} &
      \resizebox{45mm}{!}{\includegraphics{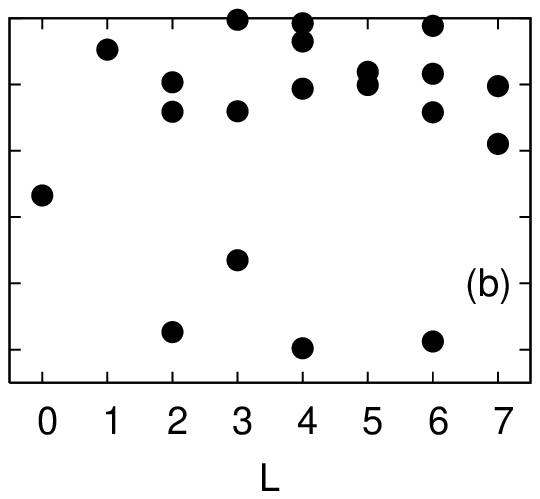}} \\
      \resizebox{45mm}{!}{\includegraphics{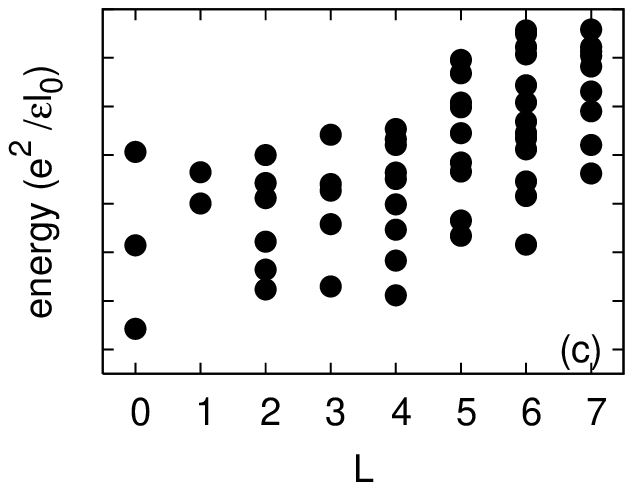}} &
      \resizebox{45mm}{!}{\includegraphics{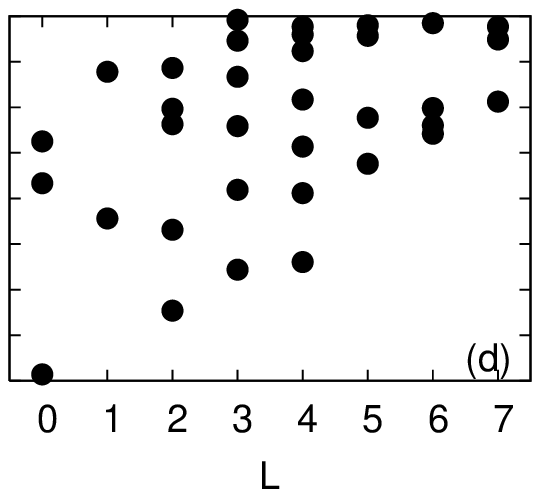}} \\
    \end{tabular}
    \caption{Low-lying spectrum of 9 electrons in the LLL
     as a function of the angular momentum
     on a sphere. 
     Top panel~: Coulomb interaction.
     (a) At
     $N_\phi=16$, the flux needed for the 3/7 CF state there is a singlet ground state
     and a branch of collective excitations. (b) At the flux needed for the candidate state
     there is no evidence of a FQHE state.
     Bottom panel weakened potential with $V_1 = 0.7V_1^{Coulomb}$~:
      (c) The 3/7 CF state is now compressible.
      (d) There is a possible new FQHE state
     with the shift required by Eq(\ref{NGwavef}).}
    \label{coulomb37}
  \end{center}
\end{figure}

\begin{figure}
    \begin{center}
      \resizebox{70mm}{!}{\includegraphics{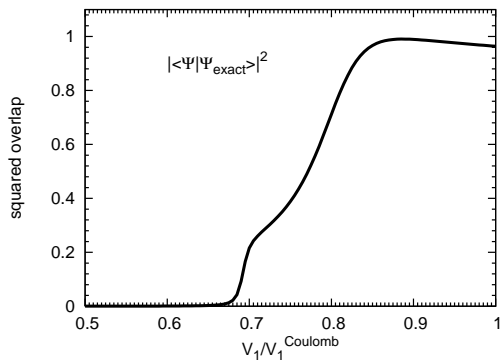}}
      \caption{The squared overlap  for N=6 electrons at $N_\phi=15$ between the candidate
        wavefunction at $\nu =3/7$
        and the exact ground state computed by varying the pseudopotential $V_1$ with respect
         to the Coulomb value.}
      \label{overlap}
    \end{center}
  \end{figure}

\begin{figure}
    \begin{center}
      \resizebox{70mm}{!}{\includegraphics{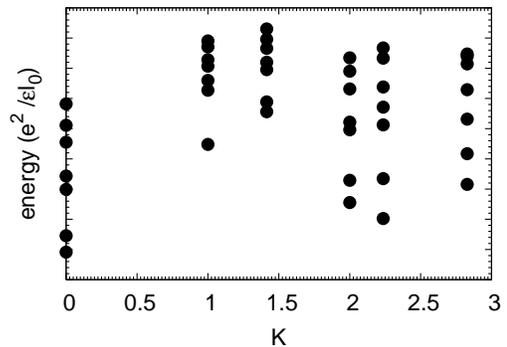}}
      \caption{Low-lying spectrum  for N=12 electrons at $N_\phi=28$ for weakened 
         pseudopotential $\delta V_1=-0.125$ in a square unit cell vs 
       pseudomomentum $K=|\mathbf{K}|$.}
      \label{T37}
    \end{center}
  \end{figure}

I now show that a similar phenomenon happens at $\nu =3/7$ in the LLL. The conventional CF
state at this filling factor is a member of the principal sequence of states. It
is realized for $N=9$ electrons at $N_\phi =16$ in the spherical geometry. There is a 
singlet ground state and a well-defined branch of neutral excitations
for $L=2,3,4,5$~: see Fig.(\ref{coulomb37}a).
 The negative-flux state Eq.(\ref{NGwavef})
requires $N_\phi=20$ for the same number of particles. At this flux for pure Coulomb interaction
there is simply a set of nearly degenerate states without evidence for an incompressible 
state~: see Fig.(\ref{coulomb37}b). If the pseudopotential $V_1$ is decreased from its Coulomb
LLL value, the CF state is quickly destroyed (Fig.(\ref{coulomb37}c)) but there is appearance of a possibly
incompressible state precisely at the special shift predicted above~: Fig.(\ref{coulomb37}d). 
There is a $L=0$ ground state and
a branch of excited states for $L=2,3,4$. To check if this state is really the new negative flux state
proposed above,  the overlap
between the candidate wavefunction for $k=3$ in Eq.(\ref{NGwavef}) and the numerically 
obtained ground state
is displayed in Fig.(\ref{overlap}) for N=6 electrons at $N_\phi=15$. Even for the pure 
Coulomb interaction
the squared overlap is 0.9641 and it rises up to 0.99054 for $V_1=0.885V_1^{Coulomb}$.
To investigate the competition between these states it is convenient to use the torus
geometry for which these is no shift. Translation symmetries of the many-body problem
may be used to construct~\cite{Haldane85} a conserved momentum $\mathbf{K}=(K_x,K_y)$
which is living in a Brillouin zone with $\overline{N}^2$ points where $\overline{N}$ is
the GCD of $N$ and $N_\phi$. The advantage of the torus geometry is that it reveals
the topological degeneracies of non-Abelian states like the Read-Rezayi parafermionic
states~\cite{Read96,Read99}. For the Bose $k=3$ state of Eq.(\ref{RR}), there is
a twofold degeneracy of the ground state which is exact only in the thermodynamic limit
(in addition to a global twofold degeneracy of the center of mass). The corresponding doublet 
of states has $\mathbf{K}=\mathbf{0}$. The standard hierarchy/composite fermion states
only have the center of mass degeneracy which is discarded in what follows.
Diagonalization of a system of $N=12$ electrons at $N_\phi =28$ in a square unit cell
shows evidence for a doublet ground state at $\mathbf{K}=\mathbf{0}$ in the neighborhood
of $\delta V_1=-0.125$~: see Fig.(\ref{T37}). The formation of the low-lying doublet appear 
before weakening of the states with nonzero wavevector, i.e. prior to an instability that 
breaks translation symmetry. Further weakening of $V_1$ leads to a compressible stripe 
phase beyond $\delta V_1=-0.15$,
where the spectrum changes abruptly when going to a rectangular unit cell (not shown)
with an aspect ratio of 0.95.

\begin{figure}
  \begin{center}

    \begin{tabular}{cc}
      \resizebox{45mm}{!}{\includegraphics{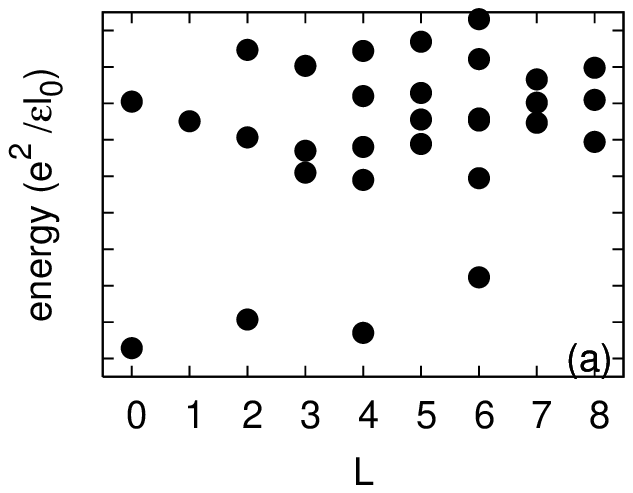}} &
      \resizebox{45mm}{!}{\includegraphics{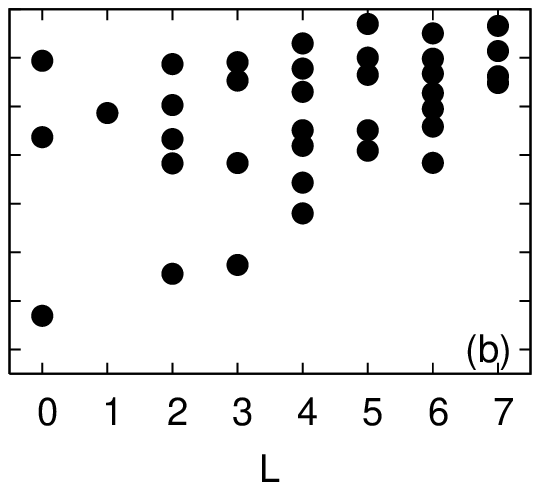}} 
    \end{tabular}
    \caption{The spectrum on a sphere for $N=6$ electrons and $N_\phi =17$
    where an unconventional state with $\nu =3/11$ may appear. (a) with pure 
    Coulomb interaction, there is a set low-lying states best viewed as
     two quasiholes on top of a $\nu =1/3$ liquid. (b) with $V_1 = 0.65 V_1^{Coulomb}$
     the $L=0$ ground state is separated from other higher-lying state
     and there is a branch of states at $L=2,3,4$}
    \label{311}
  \end{center}
\end{figure}

Another state that can be studied with present exact diagonalization techniques is $\nu =3/11$.
This fraction is obsevred experimentally and may be tentatively described by three 
filled pseudo-Landau levels of $^4$CFs. For $N=6$ the non-Abelian state with positive flux
at $k=3$ is realized on the sphere at 
$N_\phi =17$ instead of $N_\phi=21$ for the CF state. At this flux with pure Coulomb interaction, 
one finds a band
of low-lying states with $\Delta L=2$ spacing, i.e. two quasiholes states on top
of the  closeby 1/3 liquid at $N_\phi=15$~: see Fig.(\ref{311}a). If one weakens the hard-core component
of the potential, then the ground state emerges clearly from the remainder of the spectrum
with a possible branch of neutral excitations~: see Fig.(\ref{311}b).


  \begin{figure}
    \begin{center}
      \resizebox{70mm}{!}{\includegraphics{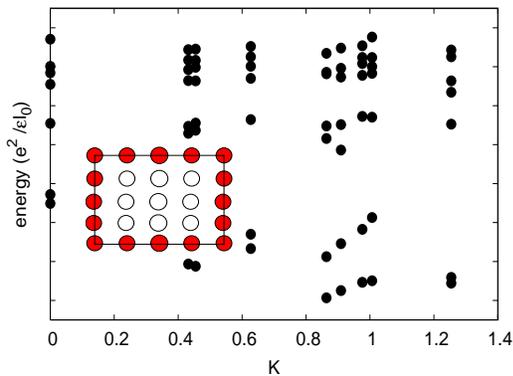}} 
      \caption{Low-lying spectrum vs. pseudomomentum $K=|\mathbf{K}|$
          for N=12 electrons at $N_\phi=32$ 
       on a torus with aspect ratio 0.95  in the LLL~:
      the sevenfold degenerate phase at $\delta V_1=+0.15$.
       Inset~: the Brillouin zone with the wavevectors (filled dots) 
     of the degenerate ground states (color online).}
      \label{T38}
    \end{center}
  \end{figure}

Finally I turn to the state at $\nu=3/8$ for which several competing states have been 
proposed~\cite{Lee01,Scarola02,Quinn03}. 
To observe
 $k=6$ clustering I use the torus
geometry with $N=12$ electrons and $N_\phi=32$. 
The phase diagram is similar for both the LLL and the N=1 LL. There is a extended phase
including the pure Coulomb point which is likely to be a kind of Fermi sea of composite particles.
There is a nondegenerate ground state ruling out broken symmetry phases
and the spectrum is very sensitive to change of the aspect ratio in the rectangular geometry.
Reducing $V_1$ leads to a transition towards a bubble phase at $\delta V_1 \approx
-0.18$ in the LLL ($\delta V_1 \approx -0.11 $ in the N=1 LL). The degenerate ground states
form a two-dimensional array in the magnetic Brillouin zone~\cite{Haldane00}
 and their number  matches
that of a bubble phase with 3 electrons per bubble~: see Fig.(\ref{T38}). This is similar to
what is observed in higher Landau levels. For larger values of $V_1$, the system enters
a phase with a seven-fold degenerate ground state for $\delta V_1\approx
+0.10$ in the LLL and $\delta V_1\approx +0.15$ in the N=1 LL~: see Fig.(\ref{T38}).
The wavevectors of the ground states lie at the boundary of the Brillouin zone and
do not define one or two dimensional ordering~: see inset of Fig.(\ref{T38})
 so a possible interpretation is a topological degeneracy. This is
what can be expected from the non-Abelian $k=6$ state of Eq.(\ref{NGwavef})
(although the wavevectors of the ground states are not known).
It is yet not possible to assess the incompressibility of this state however
it should be noted that
it is robust to changes of the aspect ratio of the torus
from unity to at least 0.85
 contrary to the 
phase in the immediate neighborhood of the Coulomb point. Also this
phase with large degeneracy survives
in the case of the pure hard-core model with $V_1$ only. This is very different from
the physics at $\nu =1/2$ in the LLL or the N=1 LL where the Pfaffian state
is confined to a narrow range of interactions.
The $k=6$ is a new candidate for the state observed at 3/8 in the LLL
as well as the 2+3/8 state. More detailed studies should await progress in understanding
the construction of such states which at the present time does not allow for overlap
calculations due to the (practical) complexity of Eq.(\ref{NGwavef}).


\begin{acknowledgments}
I thank Wei Pan for useful discussions and E. H. Rezayi for useful correspondence.
The numerical calculations have involved a computer time allocation
IDRIS-072124.
\end{acknowledgments}


\end{document}